\documentclass[longauth,letter]{aa}     
\usepackage{color}
\usepackage{graphicx}
\usepackage{rotating}
\usepackage{longtable}
\usepackage{amsmath}
\usepackage{amstext}
\usepackage{amssymb}
\usepackage{natbib}
\def   \aj {{\rm {AJ}}}

\def   \apj {{\rm {ApJ}}}

\def   \apjs {{\rm {ApJS}}}

\def   \aap {{\rm {A\&A}}}

\def   \mnras {{\rm {MNRAS}}}

\def   \apjl {{\rm {ApJL}}}

\begin{document}

\title{Substructures in the Keplerian disc around the O-type (proto)star G17.64$+$0.16 \thanks{The data presented in this article are available in electronic form
at the CDS via anonymous ftp to cdsarc.u-strasbg.fr (130.79.128.5)
or via http://cdsweb.u-strasbg.fr/cgi-bin/qcat?J/A+A/} }
\titlerunning{Substructures in the disc around G17.64$+$0.16}


\author{L.~T.~Maud\inst{1,2}\thanks{E-mail:lmaud@eso.org}  \and R.~Cesaroni\inst{3} \and M.~S.~N.~Kumar\inst{4}  \and V.~M.~Rivilla\inst{3} \and {A}.~Ginsburg\inst{5}
and P.~D.~Klaassen\inst{6} \and D.~Harsono\inst{2}  \and {\'A}.~S{\'a}nchez-Monge\inst{7} \and A.~Ahmadi\inst{8} \and V.~Allen\inst{9} \and M.~T.~Beltr{\'a}n\inst{3}
  \and H.~Beuther\inst{8} \and R.~Galv{\'a}n-Madrid\inst{10} \and C.~Goddi\inst{2,11} \and M.~G.~Hoare\inst{12} \and M.~R.~Hogerheijde\inst{2,13}
  \and K.~G.~Johnston\inst{12}  \and R.~Kuiper\inst{14} \and L.~Moscadelli\inst{3}\and T.~Peters\inst{15}
  \and L.~Testi\inst{1,3} F.~F.~S.~van der Tak\inst{16,17} \and W.~J.~de Wit\inst{18}
  }


  \authorrunning{L.~T.~Maud et al.}
  
  \institute{European Southern Observatory, Karl-Schwarzschild-Str. 2, 85748 Garching bei M{\"u}nchen, Germany \email{lmaud@eso.org}\label{inst1}
      \and Leiden Observatory, Leiden University, PO Box 9513, 2300 RA Leiden, The Netherlands\label{inst2}
    \and INAF, Osservatorio Astrofisico di Arcetri, Largo E. Fermi 5, 50125, Firenze, Italy\label{inst3}  
        \and Instituto de Astrof\'{i}sica e Ci\^{e}ncias do Espa'c{c}o, Universidade do Porto, CAUP, Rua das Estrelas, 4150–762 Porto, Portugal\label{inst4}
    \and Jansky fellow of the National Radio Astronomy Observatory, 1003 Lopezville Road, Socorro, NM 87801, USA\label{inst5}
       \and UK Astronomy Technology Centre, Royal Observatory Edinburgh, Blackford Hill, Edinburgh, EH9 3HJ, UK\label{inst6} 
    \and I. Physikalisches Institut der Universit{\"a}t zu K{\"o}ln, Z{\"u}lpicher Str. 77, 50937 K{\"o}ln, Germany\label{inst7} 
    \and Max Planck Institute for Astronomy, K{\"o}nigstuhl 17, 69117, Heidelberg, Germany\label{inst8} 
    \and NASA Postdoctoral Program Fellow, NASA Goddard Space Flight Center, Greenbelt, MD 20771, USA\label{inst9}
       \and Universidad Nacional Aut{\'o}noma de M{\'e}xico, Instituto de Radioastronom{\'i}a y Astrof{\'i}sica, Morelia, Michoac{\'a}n, 58089, Mexico\label{inst10} 
    \and Department of Astrophysics/IMAPP, Radboud University, PO Box 9010, 6500 GL Nijmegen, The Netherlands\label{inst11} 
       \and School of Physics and Astronomy, University of Leeds, Leeds, LS2 9JT, UK\label{inst12}
     \and Anton Pannekoek Institute for Astronomy, University of Amsterdam, Science Park 904, 1098 XH, Amsterdam, The Netherlands\label{inst13} 
        \and Institute of Astronomy and Astrophysics, University of T{\"u}bingen, Auf der Morgenstelle 10, 72076 T{\"u}bingen, Germany\label{inst14} 
            \and Max-Planck-Institut f{\"u}r Astrophysik, Karl-Schwarzschild-Str.1, 85748 Garching, Germany\label{inst15}  
      \and Kapteyn Astronomical Institute, University of Groningen, The Netherlands\label{inst16} 
    \and SRON, Landleven 12, 9747 AD, Groningen, The Netherlands\label{inst17} 
    \and European Southern Observatory, Al{\'o}nso de Cordova 3107, Vitacura, Casilla, 19001, Santiago de Chile, Chile\label{inst18}
   }

  \date{Received April 2019 / Accepted in A\&A June 2019}

  \abstract{We present the highest angular resolution ($\sim$20$\times$15\,mas - 44$\times$33\,au) Atacama Large Millimeter/sub-millimeter Array (ALMA) observations currently possible of the proto-O-star G17.64+0.16 in Band 6. The Cycle 5 observations with baselines out to 16\,km probes scales $<$50\,au and reveal the rotating disc around G17.64+0.16, a massive forming O-type star. The disc has a ring-like enhancement in the dust emission, especially visible as arc structures to the north and south. The Keplerian kinematics are most prominently seen in the vibrationally excited water line, H$_2$O 5$_{5,0}-$6$_{4,3}\,\nu_2=1$ ($E_u$=3461.9\,K). The mass of the central source found by modelling the Keplerian rotation is consistent with 45$\pm$10\,M$_{\odot}$. The H30$\alpha$ (231.9\,GHz) radio-recombination line and the SiO (5-4) molecular line were detected at up to the $\sim10\sigma$ level. The estimated disc mass is 0.6$-$2.6\,M$_{\odot}$ under the optically thin assumption. Analysis of the Toomre Q parameter, in the optically thin regime, indicates that the disc stability is highly dependent on temperature. The disc currently appears stable for temperatures $>$150\,K, this does not preclude that the substructures formed earlier through disc fragmentation.}
  


\maketitle

\section{Introduction}
\label{intro}

Spiral arms or rings and gaps in the discs of solar mass protostars are now common place, detectable at both IR and mm wavelengths \citep{Alma2015, Brandt2014,Andrews2016, Andrews2018,Walsh2017,Monnier2019,Boer2016}. Ring-gap structures have been explained by planets, zonal-flows, snow-lines, or dust trapping \citep{Nazari2019,Isella2018,Dipierro2015,Ruge2016,Zhang2015}, while spiral structures could be caused by interactions with planets or stellar binaries, or due to a gravitational instability of the disc itself \citep{Benisty2017,Quillen2005,Meru2017,Mayer2016}. In the context of massive star formation, there are only a handful of sources that have convincing evidence of disc rotation on sub-1000\,au scales \citep[e.g.][]{Johnston2015,Ilee2016,Ginsburg2018,Moscadelli2014,Moscadelli2019,Zapata2019}. The detection of discs and rotation can still be somewhat difficult to identify even in the few targets that have been probed at sub-100\,au resolution. Work by \citet{Beuther2019} indicate a highly fragmented star formation region in G351.77-0.54, with 12 identified structures, and tentative evidence of rotation in a few cores, while \citet{Goddi2018} observed the W51 region and also indicate a highly clustered complex environment with little, if any, evidence for stable discs.

\begin{figure*}[ht!]
\begin{center}
\includegraphics[width=1.0\textwidth]{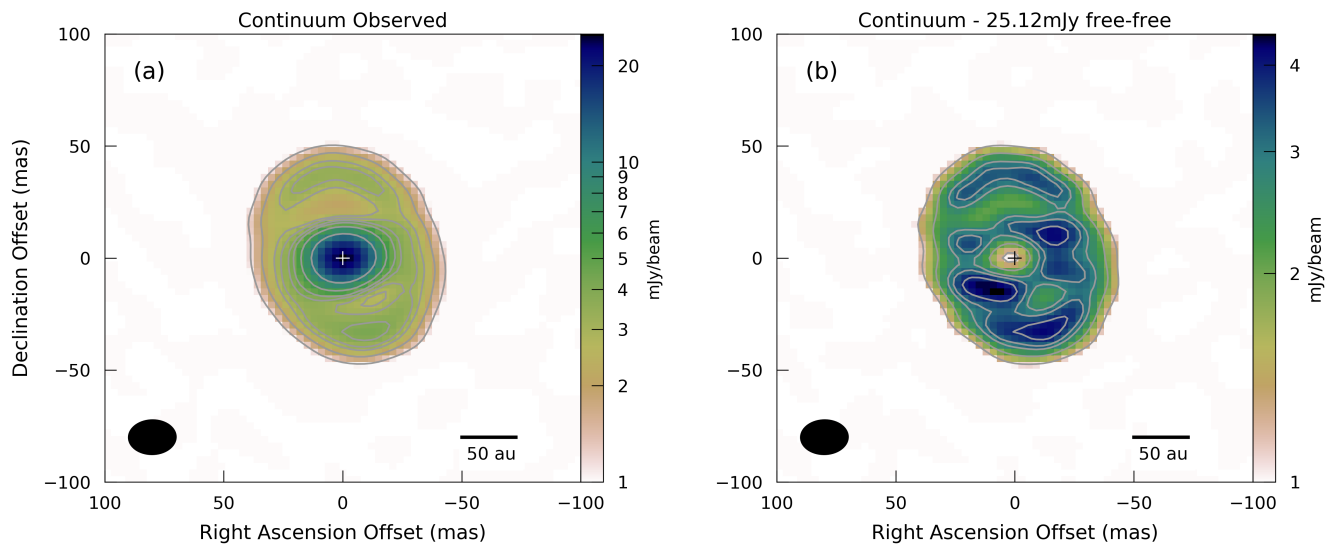}
\caption{(a)ALMA band 6 (1.3\,mm) long-baseline continuum image of G17.64 made at a resolution of 20$\times$15\,mas, PA -88.4$^{\circ}$. The enhanced emission is clearest in the north and south between 65$-$97\,au in the radial direction from G17.64. (b) As (a) but with a point source of peak flux 25.12\,mJy\,beam$^{-1}$ removed from the visibilities, representing the strongest free-free contamination in G17.64. Additional substructures now become clear. Note the change in the colour bar scale. All contours are drawn at 30, 50, 70, 80, 95, 110, 150, 250$\sigma$ of the respective images, where 1$\sigma$ = 40.4\,$\mu$Jy\,beam$^{-1}$. The beam and scale bar are indicated at the bottom. }
\label{fig:fig1} 
\end{center}
\end{figure*}

Theoretical works involving the formation of massive stars must invoke accretion discs. Simulations typically result in large-scale (500$-$1000\,au) spiral-like features or streamers that funnel the accretion flows \citep{Meyer2018,Harries2017,Krumholz2009,Peters2010b, Kuiper2011,Klassen2016, Rosen2016,Kuiper2018}, which themselves have recently been observed in a few cases \citep{Maud2017,Izquierdo2018,Liu2015,Cesaroni2014}. Multiple, or binary systems are clearly predicted by those simulations, and the substructures, either spirals, rings or fragmented discs around massive protostars should be possible to image using ALMA \citep[e.g.][]{Jankovic2019,Meyer2018}. 

Studies indicate that over 70\,\% of main-sequence OB stars \citep{Sana2012,Moe2017}, and 50\,\% of massive young stellar objects \citep{Pomohaci2019} are known to display binarity or multiplicity. In the deeply embedded star formation stages very high angular resolution ($<$100\,au scales) sub-mm observations are required to probe the natal environments. There is one recent example of a binary in a proto-O-star system where the secondary, separated by $\sim$1200\,au, is still within the disc and thought to have formed by disc fragmentation \citep{Ilee2018}. Furthermore, \citet{Zhang2019} present ALMA long-baseline observations that resolve a high-mass binary system (total-mass\,$\sim$18\,M$_{\odot}$) with a physical separation of $\sim$180\,au. 

However, rings, gaps or spiral substructures below 500\,au spatial scales have not yet been detected in discs around massive-protostars.

G17.64$+$0.16 (hereafter G17.64, also AFGL 2136, G017.6380+00.1566, CRL 2136, and IRAS 18196-1331) is a well-known massive young stellar object (MYSO) that we originally targeted with ALMA along with five other luminous O-type (proto)stars in search for evidence of discs \citep{Cesaroni2017}.  Located at 2.2\,kpc and with a bolometric luminosity of 1$\times$10$^5$\,L$_{\odot}$\footnote{Red MSX Survey: http://rms.leeds.ac.uk/cgi-bin/public/RMS\_DATABASE.cgi} \citep{Lumsden2013} G17.64 is one of the closest O-type (proto)stars. It is a bright source at near- to mid-IR wavelengths \citep[][]{Kastner1992,Holbrook1998, deWit2009,Murakawa2013} and is detected through to the cm regime \citep[e.g.][]{vandertak2000b, Menten2004,Lu2014}. It drives an outflow (position-angle $\sim$135$^{\circ}$), illuminates a reflection nebulae, and excites H$_2$O masers \citep{Menten2004}. Interferometric IR observations indicated G17.64 as a candidate compact ($<$100\,au) disc source \citep{Boley2013}. \citet{Maud2018} presented 0.2$''$ resolution ALMA data probing scales down to $\sim$400\,au. They did not resolve the continuum emission, but marginally resolved the SiO emission that is thought to trace a rotating disc and disc wind. Modelling the position-velocity profile they estimated a central source mass between 20 and 30\,M$_{\odot}$. 

In this letter we report on the observations of G17.64 using the ALMA long-baselines at Band 6. Achieving a resolution of 20$\times$15\,mas (44$\times$33\,au), ten times higher than our previous study, we now fully resolve the dust and molecular line emission and for the first time uncover clear enhanced substructures in the disc around this massive forming O-star.

\begin{figure*}
\begin{center}. 
\includegraphics[width=0.95\textwidth]{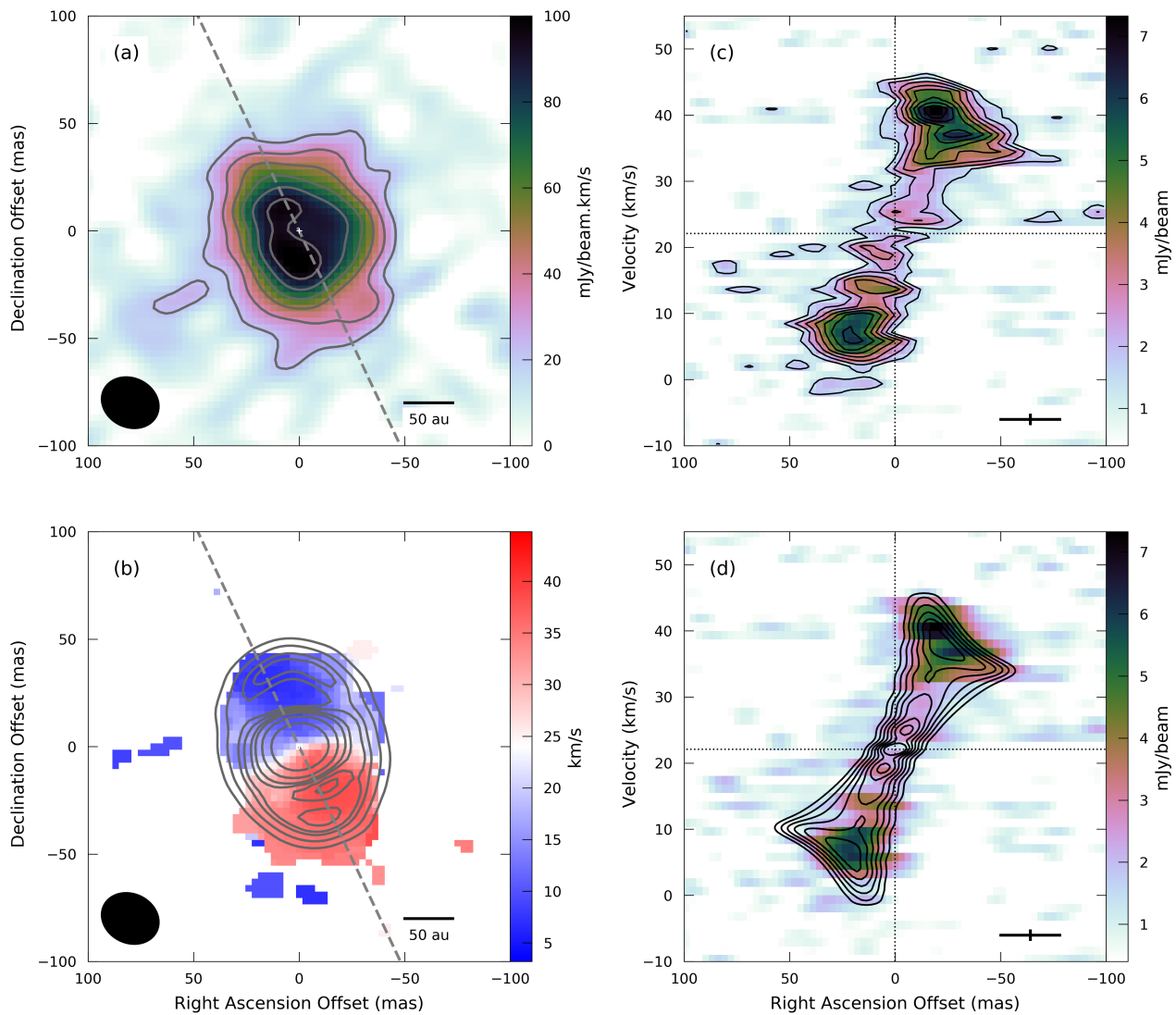}
\caption{(a) Integrated intensity (moment-0) map of the H$_2$O 5$_{5,0}-$6$_{4,3}\,\nu_2$=1(232.68670\,GHz) molecular line emission. The contours trace emission at the 3, 5, 7, 9, 11$\sigma$ levels, where 1$\sigma$ = 7.13\,mJy\,beam$^{-1}$\,km\,s$^{-1}$. (b) Velocity-weighted integrated intensity (moment-1) map overlaid with the contours of the continuum emission from Fig.\ref{fig:fig1}a in grey. Note the outer extent of the H$_2$O emission matches that of the dust continuum. (c) Position velocity diagram for the H$_2$O emission extracted from a 5\,pixel wide slice along the dashed line at PA = 25.9$^{\circ}$ as indicated in (a) and (b). The contours of the data are plotted at the 20 to 80\,\% levels in steps of 10\,\%, where 10\,\% corresponds to $\sim$1$\sigma$ (0.65\,mJy\,beam$^{-1}$). (d) As (c) but where the contours indicate the best representative model using inner and outer disc radii of 30 and 120\,au respectively, a stellar mass of 45\,M$_{\odot}$ and a disc inclination angle of 40$^{\circ}$. The spatial (28\,mas) and velocity resolution (1.3\,km\,s$^{-1}$) are indicated by the black cross to the bottom right.}
\label{fig:fig2} 
\end{center}
\end{figure*}

\section{Observations}
\label{obs}
The ALMA 12\,m observations consist of one execution block (EB) conducted during Cycle 5 on 4 October 2017 (project ID: 2017.1.00098.S - PI: Maud) in configuration C43-10, with 43 antennas. The on-source time was 30.6\,minutes. The spectral setup covered the frequency range of our previous observations \citep[see also][]{Cesaroni2017}. To provide maximal sensitivity to the dust continuum, all four spectral-windows (SPWs) were configured with the widest bandwidth of 1.875\,GHz but covered previously detected lines, e.g. SiO (5$-$4) and H30$\alpha$. The resulting velocity resolution was of the order 1.3\,km\,s$^{-1}$, except in the SPW covering SiO which was 0.8\,km\,s$^{-1}$. We also covered the H$_2$O 5$_{5,0}-$6$_{4,3}$ line in the same SPW as H30$\alpha$. The maximal angular resolution of 20$\times$15\,mas at a position-angle (PA) of -88.4$^{\circ}$ in the continuum was achieved using a robust parameter of 0.0 \citep{Briggs1995}. Data calibration used the {\sc casa} pipeline, version 5.1.1 \citep{McMullin2007}, while subsequent imaging and self-calibration used version 5.4.0. As G17.64 is comparably line weak \citep{Cesaroni2017} all line-free regions were easily identifiable and continuum subtraction was undertaken in the {\emph u,v} domain. Self-calibration was possible down to a solution time of 6\,s in phase and 54\,s for amplitude, improving the dynamic range from 405 to 640. The final continuum noise level achieved was 40.4\,$\mu$Jy\,beam$^{-1}$. For the H$_2$O images a robust value of 1.5 was used to boost surface brightness sensitivity. The resulting resolution was 28$\times$23\,mas at a PA of 65.8$^{\circ}$ and achieved a sensitivity of 0.76\,mJy\,bm$^{-1}$\,ch$^{-1}$ (1.3\,km\,s$^{-1}$). We detected SiO (5$-$4) and the H30$\alpha$ radio-recombination line which are mentioned in the appendix. During self-calibration we shifted the phase centre of G17.64 to the position of peak emission, J2000 18$^{\rm h}$22$^{\rm m}$26.3862$^{\rm s}$ $-$13$^{\circ}$30$'$11.9717$''$, to centralise our images. All imaging and self-calibration steps were performed with and without shifting the phase centre to ensure the features detected are real and not interferometric side-lobe artefacts, no notable differences were seen in any of our images.

\section{Results} 
\label{res}

\subsection{Continuum emission}
\label{contemm}

Figure \ref{fig:fig1}a shows our image of the continuum emission from G17.64 in a logarithmic colour scaling to highlight the fainter emission. The continuum dust disc is well resolved and has a clear enhanced ring-like structure that is most readily visible to the north and south between 65$-$97\,au in the radial direction. 

Fitting an ellipse to the 50$\sigma$ contour level ($\sim$2.0\,mJy\,beam$^{-1}$) we find that the disc measures 93$\times$71\,mas. Assuming a circular thin-disc where the mm-sized dust grains have settled to the mid-plane \citep[e.g.][]{Testi2014} we estimate the disc inclination as $\sim$40$\pm$4$^{\circ}$ (where 90$^{\circ}$ is edge-on). The uncertainty propagates from fitting the disc size at the 30$\sigma$ and 70$\sigma$ contours and calculation of the respective inclination angles. The peak flux is 25.12\,mJy\,beam$^{-1}$, while the enhanced arc regions peak at between 3$-$4\,mJy\,beam$^{-1}$. We recover an integrated flux of 79.14\,mJy (within a 60\,mas radius circle) which is entirely consistent with the previous ALMA observations \citep[81.3\,mJy before free-free subtraction,][]{Maud2018} and confirms that we are not resolving out any emission and can attribute the total flux entirely to the disc. Using a minimum value of 250\,m when considering the well sampled short baselines, we note our maximal recoverable scale is at least 0.64$''$, much larger than the dust disc.

In \citet{Maud2018} we reported that the radio wavelength emission accounts for between 3.57\,mJy and 29.5\,mJy when extrapolated to the mm regime. Considering the peak flux is 25.12\,mJy\,beam$^{-1}$, this places an upper limit to the free-free contamination in the case that it fully represents the mm emission. In Fig. \ref{fig:fig1}b we show the image of G17.64 after the subtraction of a point source, made in the visibility domain, using the upper limit of the free-free contamination. A point source assumption is consistent with the size (29\,mas at 43\,GHz) found by \citet{Menten2004} and the reduction in angular size with increasing frequency ($\theta\,\propto\,\nu^{-0.6}$, \citealt{Wright1975}). Note, the removal of a point-source in the visibility domain can be understood as removing a 2D gaussian with the synthesised beam parameters in the image plane. After the removal of the maximal and minimal free-free components the resulting lower limit optically thin disc mass ranges from 0.6 to 2.6\,M$_{\odot}$ following \citet{Hildebrand1983} and using dust temperatures of 50\,K, 100\,K and 150\,K consistent with \citet{Maud2018} and Section \ref{temps} (see appendix for details).

\citet{Jankovic2019} report that total removal of a central gaussian feature can be a useful technique to highlight residual substructures. It is coincidental that the necessary removal of the free-free contamination from G17.64 has a similar effect. In Fig. \ref{fig:fig1}b the brightest structure in the dust disc has a peak flux of 4.4\,mJy\,beam$^{-1}$. Although speculative, because the free-free contamination is uncertain, we tentatively reveal an enhanced inner ring or poorly resolved spiral-like structure.

\begin{figure*}[ht!]
\begin{center}
\includegraphics[width=0.95\textwidth]{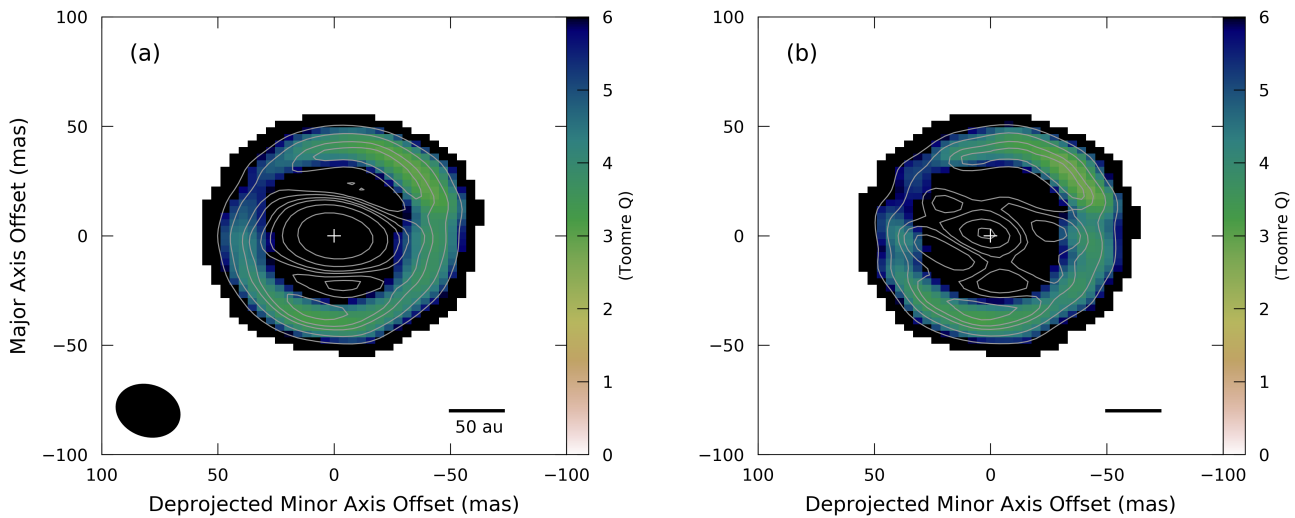} 
\caption{(a) Toomre Q map of G17.64 using the dust emission with the minimal free-free  fraction subtracted for the mass surface density (b) As (a), but where the maximal free-free contamination was removed before calculating the mass surface density. The synthesised beam is indicated to the bottom left of the left figure and a scale to the right, while the grey contours are those shown in Fig. \ref{fig:fig1}a and \ref{fig:fig1}b. Values below the 10$\sigma$ dust continuum emission level are masked out. The maps are centred (0,0) on G17.64  but are rotated by the PA  (25.9$^{\circ}$) aligning the disc major axis in the north-south direction. The disc is deprojected along the minor axis using the inclination of 40$^{\circ}$ in the image plane.} 
\label{fig:fig3} 
\end{center}
\end{figure*}

\subsection{H$_2$O}
\label{water}
Figure \ref{fig:fig2} presents the moment-0 and moment-1 maps of the H$_2$O 5$_{5,0}-$6$_{4,3}\,\nu_2=1$ line at 232.68670\,GHz ($E_u$=3461.9\,K) in panels (a) and (b) and position-velocity (PV) diagrams over plotted with contours of the data and the best representative model in panels (c) and (d). The extreme excitation energy of this line makes it particularly suited for tracing very inner regions of a molecular disc \citep{Hirota2014,Ginsburg2018}. To our sensitivity and resolution limit the H$_2$O emission appears devoid of any structure and does not extend beyond the outer radius of the disc as traced by the dust continuum emission (grey contours Fig.\ref{fig:fig2}b). In all velocity channels the peak of emission is away from the centre of the dust disc, while at the highest blue- and red-shifted velocities the H$_2$O emission peaks between the enhanced continuum dust ring and the central location of G17.64. The moment-0 contours in Figure \ref{fig:fig2}a shows how the H$_2$O emission could appear like a disc with a central hole, or a single wide ring ranging from approximately half of the beam ($\sim$30\,au) to 120\,au in radius, \citep[e.g. like the SO emission in][]{Yen2014}. The cut for the PV diagram was taken at a position angle of 25.9$^{\circ}$ using a width of 5 pixels to encompass one beam of emission (grey dashed line in Fig.\ref{fig:fig2}a,b). The PA was established by fitting a line through the centres of Gaussian fits to independent moment-0 maps of the blue- (3.2$-$21.4\,km\,s$^{-1}$) and red-shifted (22.7$-$44.8\,km\,s$^{-1}$) emission and the central position of G17.64 from the dust continuum. The PV diagram depicts the clear pattern of Keplerian rotation, particularly the red-shifted emission $>$30\,km\,s$^{-1}$ where the highest velocities peak closer towards the central source. The H$_2$O is undoubtedly tracing a rotating disc. 

\section{Discussion}

\subsection{Stellar mass}
\label{stelmass}
Based on its luminosity G17.64 is firmly positioned as an O-star \citep[e.g.][]{Vacca1996}. However, previous estimates of the stellar mass, which are essential for understanding the disc stability, were not well constrained due to poorly resolved kinematics \citep{Maud2018}. Building from our previous representative models presented in \citet{Maud2018} we match the H$_2$O PV diagram with only very minor changes. We use our `disc-only' model for the H$_2$O emission, as there is no evidence for extended structure, we fix the disc outer radius to 120\,au as per the dust and H$_2$O maps and also revise the inclination angle to 40$\pm$5$^{\circ}$. Now that the disc is fully resolved we have a measure of the system inclination which was a previously degenerate parameter \citep{Maud2018}. The inner radius is varied only between 25$-$30\,au in line with the lack of strong H$_2$O emission at the centre \citep[Fig.2a, see also][]{Maud2018}, and consistent with the potentially ionised inner disc seen in H30$\alpha$ emission (appendix). With these modifications, the modelled central mass must be increased to 45$\pm$10\,M$_{\odot}$ to best match the observed PV diagram (Fig.\ref{fig:fig2}c,d), especially the red-shifted emission at higher velocities. The mass increase compared to \citet{Maud2018} is due to the revised inclination angle as the disc is more face-on. Following \citet{Maud2018}, fitting is made by-eye matching equivalent contour levels \citep[see also][]{Ilee2018}.

\subsection{Disc stability}
\label{temps}
In discs undergoing Keplerian rotation, local regions may be unstable to axisymmetric perturbations if the Toomre parameter, Q, \citep{Toomre1964} is:

\begin{equation} 
Q = \frac{c_s \kappa}{\pi G \Sigma} < 1, 
\end{equation}
where $c_s$ is the sound speed, $\kappa$ is the epicyclic frequency - orbital velocity $\Omega$ for a Keplerian disc - G is the gravitational constant, and $\Sigma$ is the disc surface density.
Figure \ref{fig:fig3}a,b show the Toomre Q maps for G17.64 where the minimal and maximal free-free contamination is removed. A deprojection has been applied in the image plane for the minor disc axis after rotation (major axis is north-south). We follow the methods outlined in \citet{Beuther2017a} and \citet{Ahmadi2018}. Both the mass surface density, calculated from the column density \citep[Eq 2.][]{Schuller2009} multiplied by the mean molecular weight and mass of the hydrogen atom ($\mu m_{\rm H}$, where $\mu$ = 2.3), and the sound speed use a radial temperature dependence and we directly calculate the orbital velocity from Keplerian rotation. The temperature of the disc is calculated using $T(R)\propto (R/R_0)^{-0.4}$ \citep{Whitney2003} propagating from the outer radius and temperature of a 45\,M$_{\odot}$ main-sequence O-star from \citet{Hosokawa2009}, where for G17.64 we estimated, $R_0\sim$9.3\,R$_{\odot}$, $L\sim$1.9$\times$10$^{5}$\,L$_{\odot}$ and $T\sim$60000\,K. At radii $<$150\,au the disk temperature is $>$150\,K.

The ring-like enhancement between radii 65$-$97\,au has the lowest Toomre Q values, between 2 and 4. There is little difference between the data with the minimal and maximal free-free contamination removed as the lowest Q regions are away from the central peak where the free-free contamination occurs. There are no current instabilities identified in the substructures (c.f. Fig.\ref{fig:fig1}b). If we use a steeper temperature profile \citep[e.g. -0.5,][]{Brinch2010} the temperature in the disc is below 150\,K, and the overall cooler disc can become unstable Q$<$1. Additionally, in an optically thick case, suggested by the continuum average brightness temperatures (see the appendix), the surface density will increase when correcting for opacity and Q would decrease. If considering a highly flared disc, the Toomre criterion is met where Q$<$1.7 \citep{Durisen2007}, thus a proxy of the dust and gas disc scale heights and temperature estimation from radiative transfer are required to better constrain the stability. The presented Toomre Q analysis suggests that unless the dust is cold and significantly shielded the disc can remain stable against fragmentation. This does not preclude that the existing substructures may have formed from fragmentation. During formation, a more massive, cooler, and potentially largely flared disc combined with a reduced stellar mass would have yielded conditions where Q<1.7.

\subsection{Disc enhancements}
G17.64 is the first massive O-type source in which substructure has been observed in the disc. Interestingly, recent IR interferometry and models by \citet{Frost2019} suggest that the MYSO G305.20+0.21 could be a massive transition disc, with an inner edge at 60\,au, not dissimilar to the ring-enhancement in G17.64, although there are no high-resolution mm studies. Similar processes, such as radial drifts, could be at work in G17.64 as they are in low-mass sources \citep{Pinilla2018}. In this case dust grains become trapped in pressure maxima and can grow (see Sec.\ref{intro}), however, multi-wavelength observations are required to establish dust grain sizes and confirm the phenomena. Considering that most massive systems exist as binaries, it is plausible that a binary companion is the cause of the ring-like enhancements, much like the 1.8\,M$_{\odot}$ source HD\,142527 \citep{Price2018}.  We could interpret the brightest dust peak south-east of G17.64+0.16 as a binary companion, although there is no clearly separated source like in \citet{Ilee2018}. The substructures could be spirals caused by instabilities occurring before this `stable' disc phase as these would appear as arc or ring-like enhancements if not fully resolved. Higher-frequency observations may be the only way to probe 10\,au scales, provided that substructures are still visible with increasing optical depth.

\section{Conclusions} 
\label{conc}
In this letter we present ALMA long baseline observations that reveal the disc in the proto-O-star G17.64$+$0.16. Through the PV analysis of the H$_2$O emission using our disc model we confirm Keplerian rotation about
a central mass of 45$\pm$10\,M$_{\odot}$. We find that the continuum emission has a significant enhancement of dust emission in a ring-like, or possibly underlying spiral structure. The disc is found to be locally stable through a Toomre Q analysis in the optically thin case when disc temperature are $>$150\,K.

\begin{acknowledgements}
The authors thank the referee for their detailed comments that helped to improve this manuscript. MSNK acknowledges the support from Funda\c{c}\~ao para a Ci\^{e}ncia e Tecnologia
through Investigador contracts and exploratory project IF/00956/2015/CP1273/CT0002
HB and AA acknowledge support from the European Research Council under the European Community's Horizon 2020 framework program (2014-2020) via the ERC Consolidator Grant `From Cloud to Star Formation (CSF)' (project number 648505). RGM acknowledges support from UNAM-PAPIIT Programme IN104319. 
  RK acknowledges financial support via the Emmy Noether Research Group funded by the German Research Foundation (DFG) under grant no. KU 2849/3-1 and KU 2849/3-2.
  VMR thanks the funding from the European Union's Horizon 2020 research and innovation programme under the Marie Sk\l{}odowska-Curie grant agreement No 664931.
  This paper makes use of the least-squares fitting by \citet{HammelLS}.
  This paper makes use of the following ALMA data: ADS/JAO.ALMA\#2017.1.00098.S. ALMA is a partnership of ESO (representing its member states), NSF (USA) and NINS (Japan), together with NRC (Canada), NSC and ASIAA (Taiwan), and KASI (Republic of Korea), in cooperation with the Republic of Chile. The Joint ALMA Observatory is operated by ESO, AUI/NRAO and NAOJ.

\end{acknowledgements}
\bibliographystyle{aa}

\begin{thebibliography}{74}
\expandafter\ifx\csname natexlab\endcsname\relax\def\natexlab#1{#1}\fi

\bibitem[{{Ahmadi} {et~al.}(2018){Ahmadi}, {Beuther}, {Mottram}, {Bosco},
  {Linz}, {Henning}, {Winters}, {Kuiper}, {Pudritz}, {S{\'a}nchez-Monge},
  {Keto}, {Beltran}, {Bontemps}, {Cesaroni}, {Csengeri}, {Feng},
  {Galvan-Madrid}, {Johnston}, {Klaassen}, {Leurini}, {Longmore}, {Lumsden},
  {Maud}, {Menten}, {Moscadelli}, {Motte}, {Palau}, {Peters}, {Ragan},
  {Schilke}, {Urquhart}, {Wyrowski}, \& {Zinnecker}}]{Ahmadi2018}
{Ahmadi}, A., {Beuther}, H., {Mottram}, J.~C., {et~al.} 2018, \aap, 618, A46

\bibitem[{{ALMA Partnership} {et~al.}(2015){ALMA Partnership}, {Brogan},
  {P{\'e}rez}, {Hunter}, {Dent}, {Hales}, {Hills}, {Corder}, {Fomalont},
  {Vlahakis}, {Asaki}, \& et~al.}]{Alma2015}
{ALMA Partnership}, {Brogan}, C.~L., {P{\'e}rez}, L.~M., {et~al.} 2015, \apjl,
  808, L3

\bibitem[{{Almeida} {et~al.}(2017){Almeida}, {Sana}, {Taylor}, {Barb{\'a}},
  {Bonanos}, {Crowther}, {Damineli}, {de Koter}, {de Mink}, {Evans}, {Gieles},
  {Grin}, {H{\'e}nault-Brunet}, {Langer}, {Lennon}, {Lockwood}, {Ma{\'\i}z
  Apell{\'a}niz}, {Moffat}, {Neijssel}, {Norman}, {Ram{\'\i}rez-Agudelo},
  {Richardson}, {Schootemeijer}, {Shenar}, {Soszy{\'n}ski}, {Tramper}, \&
  {Vink}}]{Moe2017}
{Almeida}, L.~A., {Sana}, H., {Taylor}, W., {et~al.} 2017, \aap, 598, A84

\bibitem[{{Andrews} {et~al.}(2018){Andrews}, {Huang}, {P{\'e}rez}, {Isella},
  {Dullemond}, {Kurtovic}, {Guzm{\'a}n}, {Carpenter}, {Wilner}, {Zhang}, {Zhu},
  {Birnstiel}, {Bai}, {Benisty}, {Hughes}, {{\"O}berg}, \&
  {Ricci}}]{Andrews2018}
{Andrews}, S.~M., {Huang}, J., {P{\'e}rez}, L.~M., {et~al.} 2018, \apj, 869,
  L41

\bibitem[{{Andrews} {et~al.}(2016){Andrews}, {Wilner}, {Zhu}, {Birnstiel},
  {Carpenter}, {P{\'e}rez}, {Bai}, {{\"O}berg}, {Hughes}, {Isella}, \&
  {Ricci}}]{Andrews2016}
{Andrews}, S.~M., {Wilner}, D.~J., {Zhu}, Z., {et~al.} 2016, \apj, 820, L40

\bibitem[{{Benisty} {et~al.}(2017){Benisty}, {Stolker}, {Pohl}, {de Boer},
  {Lesur}, {Dominik}, {Dullemond}, {Langlois}, {Min}, {Wagner}, {Henning},
  {Juhasz}, {Pinilla}, {Facchini}, {Apai}, {van Boekel}, {Garufi}, {Ginski},
  {M{\'e}nard}, {Pinte}, {Quanz}, {Zurlo}, {Boccaletti}, {Bonnefoy}, {Beuzit},
  {Chauvin}, {Cudel}, {Desidera}, {Feldt}, {Fontanive}, {Gratton}, {Kasper},
  {Lagrange}, {LeCoroller}, {Mouillet}, {Mesa}, {Sissa}, {Vigan}, {Antichi},
  {Buey}, {Fusco}, {Gisler}, {Llored}, {Magnard}, {Moeller-Nilsson}, {Pragt},
  {Roelfsema}, {Sauvage}, \& {Wildi}}]{Benisty2017}
{Benisty}, M., {Stolker}, T., {Pohl}, A., {et~al.} 2017, \aap, 597, A42

\bibitem[{{Beuther} {et~al.}(2019){Beuther}, {Ahmadi}, {Mottram}, {Linz},
  {Maud}, {Henning}, {Kuiper}, {Walsh}, {Johnston}, \&
  {Longmore}}]{Beuther2019}
{Beuther}, H., {Ahmadi}, A., {Mottram}, J.~C., {et~al.} 2019, \aap, 621, A122

\bibitem[{{Beuther} {et~al.}(2017){Beuther}, {Walsh}, {Johnston}, {Henning},
  {Kuiper}, {Longmore}, \& {Walmsley}}]{Beuther2017a}
{Beuther}, H., {Walsh}, A.~J., {Johnston}, K.~G., {et~al.} 2017, \aap, 603, A10

\bibitem[{{Boley} {et~al.}(2013){Boley}, {Linz}, {van Boekel}, {Henning},
  {Feldt}, {Kaper}, {Leinert}, {M{\"u}ller}, {Pascucci}, {Robberto},
  {Stecklum}, {Waters}, \& {Zinnecker}}]{Boley2013}
{Boley}, P.~A., {Linz}, H., {van Boekel}, R., {et~al.} 2013, \aap, 558, A24

\bibitem[{{Brandt} {et~al.}(2014){Brandt}, {Kuzuhara}, {McElwain}, {Schlieder},
  {Wisniewski}, {Turner}, {Carson}, {Matsuo}, {Biller}, {Bonnefoy}, {Dressing},
  {Janson}, {Knapp}, {Moro-Mart{\'{\i}}n}, {Thalmann}, {Kudo}, {Kusakabe},
  {Hashimoto}, {Abe}, {Brandner}, {Currie}, {Egner}, {Feldt}, {Golota}, {Goto},
  {Grady}, {Guyon}, {Hayano}, {Hayashi}, {Hayashi}, {Henning}, {Hodapp},
  {Ishii}, {Iye}, {Kandori}, {Kwon}, {Mede}, {Miyama}, {Morino}, {Nishimura},
  {Pyo}, {Serabyn}, {Suenaga}, {Suto}, {Suzuki}, {Takami}, {Takahashi},
  {Takato}, {Terada}, {Tomono}, {Watanabe}, {Yamada}, {Takami}, {Usuda}, \&
  {Tamura}}]{Brandt2014}
{Brandt}, T.~D., {Kuzuhara}, M., {McElwain}, M.~W., {et~al.} 2014, \apj, 786, 1

\bibitem[{{Briggs}(1995)}]{Briggs1995}
{Briggs}, D.~S. 1995, in Bulletin of the American Astronomical Society,
  Vol.~27, American Astronomical Society Meeting Abstracts, 1444

\bibitem[{{Brinch} \& {Hogerheijde}(2010)}]{Brinch2010}
{Brinch}, C. \& {Hogerheijde}, M.~R. 2010, \aap, 523, A25

\bibitem[{{Cesaroni} {et~al.}(2014){Cesaroni}, {Galli}, {Neri}, \&
  {Walmsley}}]{Cesaroni2014}
{Cesaroni}, R., {Galli}, D., {Neri}, R., \& {Walmsley}, C.~M. 2014, \aap, 566,
  A73

\bibitem[{{Cesaroni} {et~al.}(2017){Cesaroni}, {S{\'a}nchez-Monge},
  {Beltr{\'a}n}, {Johnston}, {Maud}, {Moscadelli}, {Mottram}, {Ahmadi},
  {Allen}, {Beuther}, {Csengeri}, {Etoka}, {Fuller}, {Galli},
  {Galv{\'a}n-Madrid}, {Goddi}, {Henning}, {Hoare}, {Klaassen}, {Kuiper},
  {Kumar}, {Lumsden}, {Peters}, {Rivilla}, {Schilke}, {Testi}, {van der Tak},
  {Vig}, {Walmsley}, \& {Zinnecker}}]{Cesaroni2017}
{Cesaroni}, R., {S{\'a}nchez-Monge}, {\'A}., {Beltr{\'a}n}, M.~T., {et~al.}
  2017, \aap, 602, A59

\bibitem[{{de Boer} {et~al.}(2016){de Boer}, {Salter}, {Benisty}, {Vigan},
  {Boccaletti}, {Pinilla}, {Ginski}, {Juhasz}, {Maire}, {Messina}, {Desidera},
  {Cheetham}, {Girard}, {Wahhaj}, {Langlois}, {Bonnefoy}, {Beuzit}, {Buenzli},
  {Chauvin}, {Dominik}, {Feldt}, {Gratton}, {Hagelberg}, {Isella}, {Janson},
  {Keller}, {Lagrange}, {Lannier}, {Menard}, {Mesa}, {Mouillet}, {Mugrauer},
  {Peretti}, {Perrot}, {Sissa}, {Snik}, {Vogt}, {Zurlo}, \& {SPHERE
  Consortium}}]{Boer2016}
{de Boer}, J., {Salter}, G., {Benisty}, M., {et~al.} 2016, \aap, 595, A114

\bibitem[{{de Wit} {et~al.}(2009){de Wit}, {Hoare}, {Fujiyoshi}, {Oudmaijer},
  {Honda}, {Kataza}, {Miyata}, {Okamoto}, {Onaka}, {Sako}, \&
  {Yamashita}}]{deWit2009}
{de Wit}, W.~J., {Hoare}, M.~G., {Fujiyoshi}, T., {et~al.} 2009, \aap, 494, 157

\bibitem[{{Dipierro} {et~al.}(2015){Dipierro}, {Pinilla}, {Lodato}, \&
  {Testi}}]{Dipierro2015}
{Dipierro}, G., {Pinilla}, P., {Lodato}, G., \& {Testi}, L. 2015, \mnras, 451,
  974

\bibitem[{{Durisen} {et~al.}(2007){Durisen}, {Boss}, {Mayer}, {Nelson},
  {Quinn}, \& {Rice}}]{Durisen2007}
{Durisen}, R.~H., {Boss}, A.~P., {Mayer}, L., {et~al.} 2007, in Protostars and
  Planets V, ed. B.~{Reipurth}, D.~{Jewitt}, \& K.~{Keil}, 607

\bibitem[{{Frost} {et~al.}(2019){Frost}, {Oudmaijer}, {de Wit}, \&
  {Lumsden}}]{Frost2019}
{Frost}, A.~J., {Oudmaijer}, R.~D., {de Wit}, W.~J., \& {Lumsden}, S.~L. 2019,
  \aap, 625, A44

\bibitem[{{Ginsburg} {et~al.}(2018){Ginsburg}, {Bally}, {Goddi}, {Plambeck}, \&
  {Wright}}]{Ginsburg2018}
{Ginsburg}, A., {Bally}, J., {Goddi}, C., {Plambeck}, R., \& {Wright}, M. 2018,
  \apj, 860, 119

\bibitem[{{Goddi} {et~al.}(2018){Goddi}, {Ginsburg}, {Maud}, {Zhang}, \&
  {Zapata}}]{Goddi2018}
{Goddi}, C., {Ginsburg}, A., {Maud}, L., {Zhang}, Q., \& {Zapata}, L. 2018,
  ArXiv e-prints [\eprint[arXiv]{1805.05364}]

\bibitem[{{Hammel} \& {Sullivan-Molina}(2019)}]{HammelLS}
{Hammel}, B. \& {Sullivan-Molina}, N. 2019,
  {bdhammel/least-squares-ellipse-fitting: Initial release}

\bibitem[{{Harries} {et~al.}(2017){Harries}, {Douglas}, \& {Ali}}]{Harries2017}
{Harries}, T.~J., {Douglas}, T.~A., \& {Ali}, A. 2017, \mnras, 471, 4111

\bibitem[{{Hildebrand}(1983)}]{Hildebrand1983}
{Hildebrand}, R.~H. 1983, Quarterly Journal of the Royal Astronomical Society,
  24, 267

\bibitem[{{Hirota} {et~al.}(2014){Hirota}, {Kim}, {Kurono}, \&
  {Honma}}]{Hirota2014}
{Hirota}, T., {Kim}, M.~K., {Kurono}, Y., \& {Honma}, M. 2014, \apjl, 782, L28

\bibitem[{{Holbrook} \& {Temi}(1998)}]{Holbrook1998}
{Holbrook}, J.~C. \& {Temi}, P. 1998, \apj, 496, 280

\bibitem[{{Hosokawa} \& {Omukai}(2009)}]{Hosokawa2009}
{Hosokawa}, T. \& {Omukai}, K. 2009, \apj, 691, 823

\bibitem[{{Ilee} {et~al.}(2018){Ilee}, {Cyganowski}, {Brogan}, {Hunter},
  {Forgan}, {Haworth}, {Clarke}, \& {Harries}}]{Ilee2018}
{Ilee}, J.~D., {Cyganowski}, C.~J., {Brogan}, C.~L., {et~al.} 2018, \apj, 869,
  L24

\bibitem[{{Ilee} {et~al.}(2016){Ilee}, {Cyganowski}, {Nazari}, {Hunter},
  {Brogan}, {Forgan}, \& {Zhang}}]{Ilee2016}
{Ilee}, J.~D., {Cyganowski}, C.~J., {Nazari}, P., {et~al.} 2016, \mnras, 462,
  4386

\bibitem[{{Isella} \& {Turner}(2018)}]{Isella2018}
{Isella}, A. \& {Turner}, N.~J. 2018, \apj, 860, 27

\bibitem[{{Izquierdo} {et~al.}(2018){Izquierdo}, {Galv{\'a}n-Madrid}, {Maud},
  {Hoare}, {Johnston}, {Keto}, {Zhang}, \& {de Wit}}]{Izquierdo2018}
{Izquierdo}, A.~F., {Galv{\'a}n-Madrid}, R., {Maud}, L.~T., {et~al.} 2018,
  \mnras, 478, 2505

\bibitem[{{Jankovic} {et~al.}(2019){Jankovic}, {Haworth}, {Ilee}, {Forgan},
  {Cyganowski}, {Walsh}, {Brogan}, {Hunter}, \& {Mohanty}}]{Jankovic2019}
{Jankovic}, M.~R., {Haworth}, T.~J., {Ilee}, J.~D., {et~al.} 2019, \mnras, 482,
  4673

\bibitem[{{Johnston} {et~al.}(2015){Johnston}, {Robitaille}, {Beuther}, {Linz},
  {Boley}, {Kuiper}, {Keto}, {Hoare}, \& {van Boekel}}]{Johnston2015}
{Johnston}, K.~G., {Robitaille}, T.~P., {Beuther}, H., {et~al.} 2015, \apjl,
  813, L19

\bibitem[{{Kastner} {et~al.}(1992){Kastner}, {Weintraub}, \&
  {Aspin}}]{Kastner1992}
{Kastner}, J.~H., {Weintraub}, D.~A., \& {Aspin}, C. 1992, \apj, 389, 357

\bibitem[{{Klassen} {et~al.}(2016){Klassen}, {Pudritz}, {Kuiper}, {Peters}, \&
  {Banerjee}}]{Klassen2016}
{Klassen}, M., {Pudritz}, R.~E., {Kuiper}, R., {Peters}, T., \& {Banerjee}, R.
  2016, \apj, 823, 28

\bibitem[{{Krumholz} {et~al.}(2009){Krumholz}, {Klein}, {McKee}, {Offner}, \&
  {Cunningham}}]{Krumholz2009}
{Krumholz}, M.~R., {Klein}, R.~I., {McKee}, C.~F., {Offner}, S.~S.~R., \&
  {Cunningham}, A.~J. 2009, Science, 323, 754

\bibitem[{{Kuiper} \& {Hosokawa}(2018)}]{Kuiper2018}
{Kuiper}, R. \& {Hosokawa}, T. 2018, \aap, 616, A101

\bibitem[{{Kuiper} {et~al.}(2011){Kuiper}, {Klahr}, {Beuther}, \&
  {Henning}}]{Kuiper2011}
{Kuiper}, R., {Klahr}, H., {Beuther}, H., \& {Henning}, T. 2011, \apj, 732, 20

\bibitem[{{Liu} {et~al.}(2015){Liu}, {Galv{\'a}n-Madrid}, {Jim{\'e}nez-Serra},
  {Rom{\'a}n-Z{\'u}{\~n}iga}, {Zhang}, {Li}, \& {Chen}}]{Liu2015}
{Liu}, H.~B., {Galv{\'a}n-Madrid}, R., {Jim{\'e}nez-Serra}, I., {et~al.} 2015,
  \apj, 804, 37

\bibitem[{{Lu} {et~al.}(2014){Lu}, {Zhang}, {Liu}, {Wang}, \& {Gu}}]{Lu2014}
{Lu}, X., {Zhang}, Q., {Liu}, H.~B., {Wang}, J., \& {Gu}, Q. 2014, \apj, 790,
  84

\bibitem[{{Lumsden} {et~al.}(2013){Lumsden}, {Hoare}, {Urquhart}, {Oudmaijer},
  {Davies}, {Mottram}, {Cooper}, \& {Moore}}]{Lumsden2013}
{Lumsden}, S.~L., {Hoare}, M.~G., {Urquhart}, J.~S., {et~al.} 2013, \apjs, 208,
  11

\bibitem[{{Maud} {et~al.}(2018){Maud}, {Cesaroni}, {Kumar}, {van der Tak},
  {Allen}, {Hoare}, {Klaassen}, {Harsono}, {Hogerheijde}, {S{\'a}nchez-Monge},
  {Schilke}, {Ahmadi}, {Beltr{\'a}n}, {Beuther}, {Csengeri}, {Etoka}, {Fuller},
  {Galv{\'a}n-Madrid}, {Goddi}, {Henning}, {Johnston}, {Kuiper}, {Lumsden},
  {Moscadelli}, {Mottram}, {Peters}, {Rivilla}, {Testi}, {Vig}, {de Wit}, \&
  {Zinnecker}}]{Maud2018}
{Maud}, L.~T., {Cesaroni}, R., {Kumar}, M.~S.~N., {et~al.} 2018, \aap, 620, A31

\bibitem[{{Maud} {et~al.}(2017){Maud}, {Hoare}, {Galv{\'a}n-Madrid}, {Zhang},
  {de Wit}, {Keto}, {Johnston}, \& {Pineda}}]{Maud2017}
{Maud}, L.~T., {Hoare}, M.~G., {Galv{\'a}n-Madrid}, R., {et~al.} 2017, \mnras,
  467, L120

\bibitem[{{Mayer} {et~al.}(2016){Mayer}, {Peters}, {Pineda}, {Wadsley}, \&
  {Rogers}}]{Mayer2016}
{Mayer}, L., {Peters}, T., {Pineda}, J.~E., {Wadsley}, J., \& {Rogers}, P.
  2016, \apj, 823, L36

\bibitem[{{McMullin} {et~al.}(2007){McMullin}, {Waters}, {Schiebel}, {Young},
  \& {Golap}}]{McMullin2007}
{McMullin}, J.~P., {Waters}, B., {Schiebel}, D., {Young}, W., \& {Golap}, K.
  2007, in Astronomical Society of the Pacific Conference Series, Vol. 376,
  Astronomical Data Analysis Software and Systems XVI, ed. R.~A. {Shaw},
  F.~{Hill}, \& D.~J. {Bell}, 127

\bibitem[{{Menten} \& {van der Tak}(2004)}]{Menten2004}
{Menten}, K.~M. \& {van der Tak}, F.~F.~S. 2004, \aap, 414, 289

\bibitem[{{Meru} {et~al.}(2017){Meru}, {Juh{\'a}sz}, {Ilee}, {Clarke},
  {Rosotti}, \& {Booth}}]{Meru2017}
{Meru}, F., {Juh{\'a}sz}, A., {Ilee}, J.~D., {et~al.} 2017, \apj, 839, L24

\bibitem[{{Meyer} {et~al.}(2018){Meyer}, {Kuiper}, {Kley}, {Johnston}, \&
  {Vorobyov}}]{Meyer2018}
{Meyer}, D.~M.~A., {Kuiper}, R., {Kley}, W., {Johnston}, K.~G., \& {Vorobyov},
  E. 2018, \mnras, 473, 3615

\bibitem[{{Monnier} {et~al.}(2019){Monnier}, {Harries}, {Bae}, {Setterholm},
  {Laws}, {Aarnio}, {Adams}, {Andrews}, {Calvet}, {Espaillat}, {Hartmann},
  {Kraus}, {McClure}, {Miller}, {Oppenheimer}, {Wilner}, \&
  {Zhu}}]{Monnier2019}
{Monnier}, J.~D., {Harries}, T.~J., {Bae}, J., {et~al.} 2019, \apj, 872, 122

\bibitem[{{Moscadelli} \& {Goddi}(2014)}]{Moscadelli2014}
{Moscadelli}, L. \& {Goddi}, C. 2014, \aap, 566, A150

\bibitem[{{Moscadelli} {et~al.}(2019){Moscadelli}, {Sanna}, {Cesaroni},
  {Rivilla}, {Goddi}, \& {Rygl}}]{Moscadelli2019}
{Moscadelli}, L., {Sanna}, A., {Cesaroni}, R., {et~al.} 2019, \aap, 622, A206

\bibitem[{{Murakawa} {et~al.}(2013){Murakawa}, {Lumsden}, {Oudmaijer},
  {Davies}, {Wheelwright}, {Hoare}, \& {Ilee}}]{Murakawa2013}
{Murakawa}, K., {Lumsden}, S.~L., {Oudmaijer}, R.~D., {et~al.} 2013, \mnras,
  436, 511

\bibitem[{{Nazari} {et~al.}(2019){Nazari}, {Booth}, {Clarke}, {Rosotti},
  {Tazzari}, {Juhasz}, \& {Meru}}]{Nazari2019}
{Nazari}, P., {Booth}, R.~A., {Clarke}, C.~J., {et~al.} 2019, \mnras, 485, 5914

\bibitem[{{Ossenkopf} \& {Henning}(1994)}]{Ossenkopf1994}
{Ossenkopf}, V. \& {Henning}, T. 1994, \aap, 291, 943

\bibitem[{{Peters} {et~al.}(2010){Peters}, {Mac Low}, {Banerjee}, {Klessen}, \&
  {Dullemond}}]{Peters2010b}
{Peters}, T., {Mac Low}, M.-M., {Banerjee}, R., {Klessen}, R.~S., \&
  {Dullemond}, C.~P. 2010, \apj, 719, 831

\bibitem[{{Pinilla} {et~al.}(2018){Pinilla}, {Tazzari}, {Pascucci}, {Youdin},
  {Garufi}, {Manara}, {Testi}, {van der Plas}, {Barenfeld}, {Canovas}, {Cox},
  {Hendler}, {P{\'e}rez}, \& {van der Marel}}]{Pinilla2018}
{Pinilla}, P., {Tazzari}, M., {Pascucci}, I., {et~al.} 2018, \apj, 859, 32

\bibitem[{{Pomohaci} {et~al.}(2019){Pomohaci}, {Oudmaijer}, \&
  {Goodwin}}]{Pomohaci2019}
{Pomohaci}, R., {Oudmaijer}, R.~D., \& {Goodwin}, S.~P. 2019, \mnras, 484, 226

\bibitem[{{Price} {et~al.}(2018){Price}, {Cuello}, {Pinte}, {Mentiplay},
  {Casassus}, {Christiaens}, {Kennedy}, {Cuadra}, {Sebastian Perez}, {Marino},
  {Armitage}, {Zurlo}, {Juhasz}, {Ragusa}, {Laibe}, \& {Lodato}}]{Price2018}
{Price}, D.~J., {Cuello}, N., {Pinte}, C., {et~al.} 2018, \mnras, 477, 1270

\bibitem[{{Quillen} {et~al.}(2005){Quillen}, {Varni{\`e}re}, {Minchev}, \&
  {Frank}}]{Quillen2005}
{Quillen}, A.~C., {Varni{\`e}re}, P., {Minchev}, I., \& {Frank}, A. 2005, \aj,
  129, 2481

\bibitem[{{Rosen} {et~al.}(2016){Rosen}, {Krumholz}, {McKee}, \&
  {Klein}}]{Rosen2016}
{Rosen}, A.~L., {Krumholz}, M.~R., {McKee}, C.~F., \& {Klein}, R.~I. 2016,
  \mnras, 463, 2553

\bibitem[{{Ruge} {et~al.}(2016){Ruge}, {Flock}, {Wolf}, {Dzyurkevich},
  {Fromang}, {Henning}, {Klahr}, \& {Meheut}}]{Ruge2016}
{Ruge}, J.~P., {Flock}, M., {Wolf}, S., {et~al.} 2016, \aap, 590, A17

\bibitem[{{Sana} {et~al.}(2012){Sana}, {de Mink}, {de Koter}, {Langer},
  {Evans}, {Gieles}, {Gosset}, {Izzard}, {Le Bouquin}, \&
  {Schneider}}]{Sana2012}
{Sana}, H., {de Mink}, S.~E., {de Koter}, A., {et~al.} 2012, Science, 337, 444

\bibitem[{{Schuller} {et~al.}(2009){Schuller}, {Menten}, {Contreras},
  {Wyrowski}, {Schilke}, {Bronfman}, {Henning}, {Walmsley}, {Beuther},
  {Bontemps}, {Cesaroni}, {Deharveng}, {Garay}, {Herpin}, {Lefloch}, {Linz},
  {Mardones}, {Minier}, {Molinari}, {Motte}, {Nyman}, {Reveret}, {Risacher},
  {Russeil}, {Schneider}, {Testi}, {Troost}, {Vasyunina}, {Wienen}, {Zavagno},
  {Kovacs}, {Kreysa}, {Siringo}, \& {Wei{\ss}}}]{Schuller2009}
{Schuller}, F., {Menten}, K.~M., {Contreras}, Y., {et~al.} 2009, \aap, 504, 415

\bibitem[{{Testi} {et~al.}(2014){Testi}, {Birnstiel}, {Ricci}, {Andrews},
  {Blum}, {Carpenter}, {Dominik}, {Isella}, {Natta}, {Williams}, \&
  {Wilner}}]{Testi2014}
{Testi}, L., {Birnstiel}, T., {Ricci}, L., {et~al.} 2014, in Protostars and
  Planets VI, ed. H.~{Beuther}, R.~S. {Klessen}, C.~P. {Dullemond}, \&
  T.~{Henning}, 339

\bibitem[{{Toomre}(1964)}]{Toomre1964}
{Toomre}, A. 1964, \apj, 139, 1217

\bibitem[{{Vacca} {et~al.}(1996){Vacca}, {Garmany}, \& {Shull}}]{Vacca1996}
{Vacca}, W.~D., {Garmany}, C.~D., \& {Shull}, J.~M. 1996, \apj, 460, 914

\bibitem[{{van der Tak} {et~al.}(2000){van der Tak}, {van Dishoeck}, {Evans},
  \& {Blake}}]{vandertak2000b}
{van der Tak}, F.~F.~S., {van Dishoeck}, E.~F., {Evans}, II, N.~J., \& {Blake},
  G.~A. 2000, \apj, 537, 283

\bibitem[{{Walsh} {et~al.}(2017){Walsh}, {Daley}, {Facchini}, \&
  {Juh{\'a}sz}}]{Walsh2017}
{Walsh}, C., {Daley}, C., {Facchini}, S., \& {Juh{\'a}sz}, A. 2017, \aap, 607,
  A114

\bibitem[{{Whitney} {et~al.}(2003){Whitney}, {Wood}, {Bjorkman}, \&
  {Wolff}}]{Whitney2003}
{Whitney}, B.~A., {Wood}, K., {Bjorkman}, J.~E., \& {Wolff}, M.~J. 2003, \apj,
  591, 1049

\bibitem[{{Wright} \& {Barlow}(1975)}]{Wright1975}
{Wright}, A.~E. \& {Barlow}, M.~J. 1975, \mnras, 170, 41

\bibitem[{{Yen} {et~al.}(2014){Yen}, {Takakuwa}, {Ohashi}, {Aikawa}, {Aso},
  {Koyamatsu}, {Machida}, {Saigo}, {Saito}, {Tomida}, \& {Tomisaka}}]{Yen2014}
{Yen}, H.-W., {Takakuwa}, S., {Ohashi}, N., {et~al.} 2014, \apj, 793, 1

\bibitem[{{Zapata} {et~al.}(2019){Zapata}, {Garay}, {Palau}, {Rodr{\'\i}guez},
  {Fern{\'a}ndez-L{\'o}pez}, {Estalella}, \& {Guzm{\'a}n}}]{Zapata2019}
{Zapata}, L.~A., {Garay}, G., {Palau}, A., {et~al.} 2019, \apj, 872, 176

\bibitem[{{Zhang} {et~al.}(2015){Zhang}, {Blake}, \& {Bergin}}]{Zhang2015}
{Zhang}, K., {Blake}, G.~A., \& {Bergin}, E.~A. 2015, \apj, 806, L7

\bibitem[{{Zhang} {et~al.}(2019){Zhang}, {Tan}, {Tanaka}, {De Buizer}, {Liu},
  {Beltr{\'a}n}, {Kratter}, {Mardones}, \& {Garay}}]{Zhang2019}
{Zhang}, Y., {Tan}, J.~C., {Tanaka}, K. E.~I., {et~al.} 2019, Nature Astronomy,
  224

\end{thebibliography}


\appendix
\section{Disc mass and other lines}
\label{AppendixA}

\begin{figure*}[ht!]
\begin{center}
\includegraphics[width=0.95\textwidth]{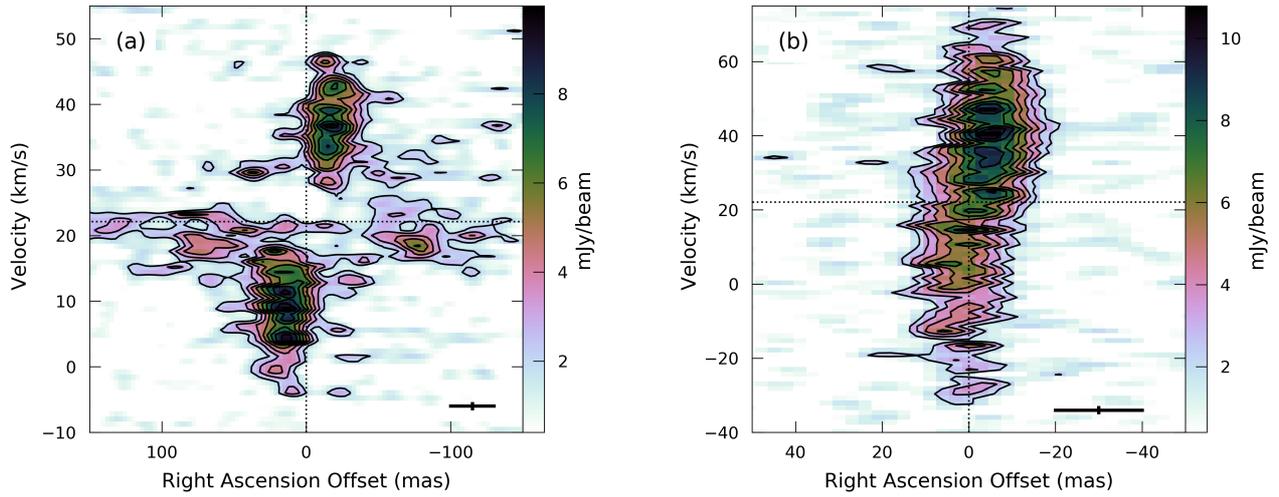}
\caption{(a) Position velocity diagram for SiO (5-4) emission taken using the same cut PA and width as for the H$_2$O presented in Fig.\ref{fig:fig2}b, a 5\,pixel wide slice along the dashed line at PA = 25.9$^{\circ}$. The resolution is shown to the bottom right as the beam major axis (30\,mas) and the velocity resolution (0.8\,km\,s$^{-1}$). Note the inner structure from $-$50 to $+$50\,mas overlaps with that of the H$_2$O emission, although we see emission extend past $\pm$50\,mas (110\,au) at low velocities \citep[c.f.][]{Maud2018}. (b) As (a) but showing the position velocity diagram for the H30$\alpha$ radio-recombination line. The resolution is shown at the bottom right, representing the 20\,mas beam major axis and a 1.3\,km\,s$^{-1}$ velocity resolution. Although tentative there appears to be a very slight shift in position of the H30$\alpha$ emission in the same rotation sense as the disc. It is possible that this is tracing very hot rotating inner material, inward of the molecule line emission in the disc. In both panels the black contours of the data are plotted at the 20 to 80\,\% levels in steps of 10\,\%. Note the axis scales and velocity range are different.}
\label{fig:figA1} 
\end{center}
\end{figure*}

\subsection*{Disc Mass}
The disc mass is estimated in the standard fashion :
\begin{equation} 
M = \frac{g S_\nu d^2}{\kappa_\nu B_\nu(T_d)}, 
\end{equation}

where $S_\nu$ is the source flux, g is the gas-to-dust ratio = 100, $B_\nu(T_d)$ is the Planck function for a temperature, $T_d$, $d$ is the source distance, and $\kappa_\nu$ is the dust opacity coefficient.
At 1.3\,mm (220\,GHz) we use $\kappa_\nu$ = 1.0 cm$^2$\,g$^{-1}$ as suggested for dust with thin ice mantles at densities of 10$^6-$10$^8$\,cm$^{-3}$ \citep{Ossenkopf1994}. As noted
we use temperatures, $T_d$, between 50-150\,K. Even when using temperatures lower than the disc average continuum brightness temperature, 193\,K and 138\,K dependent on the free-free subtraction, or those found by scaling from the source temperature (where the inner radii are slightly hotter) the disc mass should be considered as a lower limit if we are in the optically thick regime. An optical depth of $\tau$=1 would account for a disc mass increase by a factor of 1.6, to between 1.0 and 4.2\,M$_{\odot}$

\subsection*{Other Lines} 
\label{other}
We detected the previously known, strong SiO (5$-$4) molecular line emission and the H30$\alpha$ radio-recombination line. The PV diagrams for these lines are shown in Fig.\ref{fig:figA1}. The SiO also well traces the disc structure following the H$_2$O emission at disc radii of $<$120\,au, however, at low velocities closer to the source V$_{\rm LSR}$ the SiO is significantly more extended reaching out to at least 250\,au in radius before the surface brightness sensitivity is too low. The map was made with robust 1.5, the resolution is 30$\times$24\,mas at a PA of 64.9$^{\circ}$ and the sensitivity is 1.02\,mJy\,beam$^{-1}$ per 0.8\,km\,s$^{-1}$ channel. The extended emission was seen previously in our lower-resolution ALMA observations \citep{Maud2018} and is thought to be tracing outflowing material in a disc wind. 

The H30$\alpha$ emission appears unresolved per channel in the image cube and traces only the very central region of G17.64, although there is a marginal shift in spatial position from blue- to red-shifted velocities. We imaged using a robust 0.0 to provide the highest resolution (20$\times$18\,mas, PA=-88.4$^{\circ}$). The resulting sensitivity is 1.11\,mJy\,beam$^{-1}$ per 1.3\,km\,s$^{-1}$ channel. In the PV diagram, Fig.\ref{fig:figA1}b, the slight shift in position with change in velocity is clearer, and is in the same rotation sense as the H$_2$O and SiO emission. The H30$\alpha$ could be tracing a hot inner rotating structure inward of the molecular line region. We note that the total spectral coverage of our data are, SPW0: 216.746-218.621\,GHz, SPW1: 218.854-220.746\,GHz, SPW2: 230.938-232.812\,GHz, SPW3: 233.048-234.94\,GHz.

\end{document}